# Valley Degree of Freedom in Two-Dimensional van der Waals Materials

**Ashish Soni [1,2] and Suman Kalyan Pal [1,2]***

[1] School of Basic Sciences, Indian Institute of Technology Mandi, Kamand, Mandi 175005 Himachal Pradesh, India

[2] Advanced Materials Research Centre, Indian Institute of Technology Mandi, Kamand, Mandi 175005 Himachal Pradesh, India

* Email: suman@iitmandi.ac.in

**Abstract**

Layered materials can possess valleys that are indistinguishable from one another except for the momentum. These valleys are individually addressable in momentum space at the K and K' points in the first Brillouin zone. Such valley addressability opens up the possibility of utilizing the momentum state of quasi-particles as a completely new paradigm in quantum and classical information processing. This review focuses on the physics behind valley polarization and talks about carriers of valley degree of freedom (VDF) in layered materials. Then we provide a detailed survey of simple spectroscopic techniques commonly utilized to identify and manipulate valley polarization in van der Waals layered materials. Finally, we conclude with the recent developments towards the manipulation of VDF for device application and associated challenges.

## 1. Introduction

Two-dimensional materials are members of a family of layered crystal structures where atoms are connected through strong intra-layer bonds but with weak inter-layer van der Waals coupling. Graphene, an atomically thin carbon layer is the first and well-known 2D material[1, 2] It exhibits 2D Dirac fermion-like interesting features (e.g., integer and fractional quantum Hall effects)[3, 4] and ballistic conduction of charge carriers[1, 5] However, graphene is not ideal for electronic applications because of its zero bandgap semi-metallic nature. Inspired by the fascinating properties and applications of graphene, researchers developed a plethora of 2D materials. The emerging 2D materials beyond graphene include transition metal dichalcogenides (TMDCs),[6] transition metal monochalcogenides (TMMCs),[7] hexagonal boron nitride (h-BN),[8] black phosphorous (BP),[9] and graphitic carbon nitride (g-$C_3N_4$).[10] Because of extraordinary electronic properties, these novel 2D materials have shown their potential applications for diverse technologies (e.g., sensors, LEDs, FETs, catalysis, biomedicine, and environmental science)[11-16]

In a crystalline solid, a local energy minimum in the conduction band (CB) or local energy maximum in the valence band (VB) in the momentum space is known as a valley. In addition to charge and spin, the carrier is assigned with valley degree of freedom, which indicates the valley that the carrier occupies. Electrons, holes, or excitons can populate the valleys to store and carry information and form the basis for the so-called valleytronics[17-19] While electronics utilize the charge of carriers and spintronics exploits carrier spin, in valleytronics, a valley pseudospin is a quasi-particle that forms the bases of possible new technology.

Few conventional semiconductors (e.g., silicon, aluminum arsenide, bismuth) have multiple valleys in the conduction band (Fig 1a).[20] In silicon (Si), six equivalent valleys lie along the $\Delta$-direction and near to the zone boundary (X). In the recent past, efforts have been made to manipulate the VDF in traditional semiconductors.[18] Valley-polarized current is generated and detected in bulk diamond.[21] The charge conductivity of bismuth can be controlled by manipulating the polarization of Dirac valleys using a magnetic field.[22] The valley degeneracy in Si-based systems is lifted by introducing a valley splitting, which can be tuned by controlling the applied electrostatic potential.[23, 24] Despite few encouraging results, the difficulty with these materials is to maintain the VDF by using simple external agents like the electric field, magnetic field, light, strain etc.

Two-dimensional materials may intrinsically possess or be tailored to produce valley contrasting properties that can be exploited in valleytronics. In 2D-dimensional materials of hexagonal honeycomb structures (e.g., graphene, TMDCs), two valleys are present at the +K and –K points at the edges of the Brillouin zone (Fig 1b).[25] Among the 2D materials, valley contrasting physics was first detected and manipulated in graphene.[2, 26, 27] Graphene exhibits contrasted circular dichroism in different k-space regions because of inversion symmetry breaking.[28, 29] Therefore, it obeys a valley contrasting optical selection rule, which can be exploited for optoelectronics applications where light polarization information can be converted into electronic information.[26, 30, 31] In 2010, Mak et al. proposed a new direct bandgap semiconductor; 2D $MoS_2$ that exhibits very high luminescence quantum efficiency compared with the bulk material.[32, 33] Polarisation can be achieved and manipulated in monolayer $MoS_2$ through the optical pumping of circularly polarised light.[34-36] Valley selective photoluminescence (PL) have been detected in many other TMDCs (e.g., $MoSe_2$, $WS_2$, $WSe_2$).[37, 38] These 2D TMDCs show valley-dependent optical selection rule by virtue of inversion symmetry breaking due to their non-centrosymmetric hexagonally crystal lattice.

The practical realization of valleytronics devices is highly dependent on our understanding of VDF in 2D materials including manipulation of valley pseudospins. Here, we present the necessary valley physics to understand the origin of valley contrasted properties in 2D materials. The details about the carriers of valley pseudospins in layered materials will be discussed. This review will attempt to summarize the techniques currently used for investigating the VDF in 2D systems. Finally, we explore the available ways of manipulating valley polarization in 2D hexagonal materials.

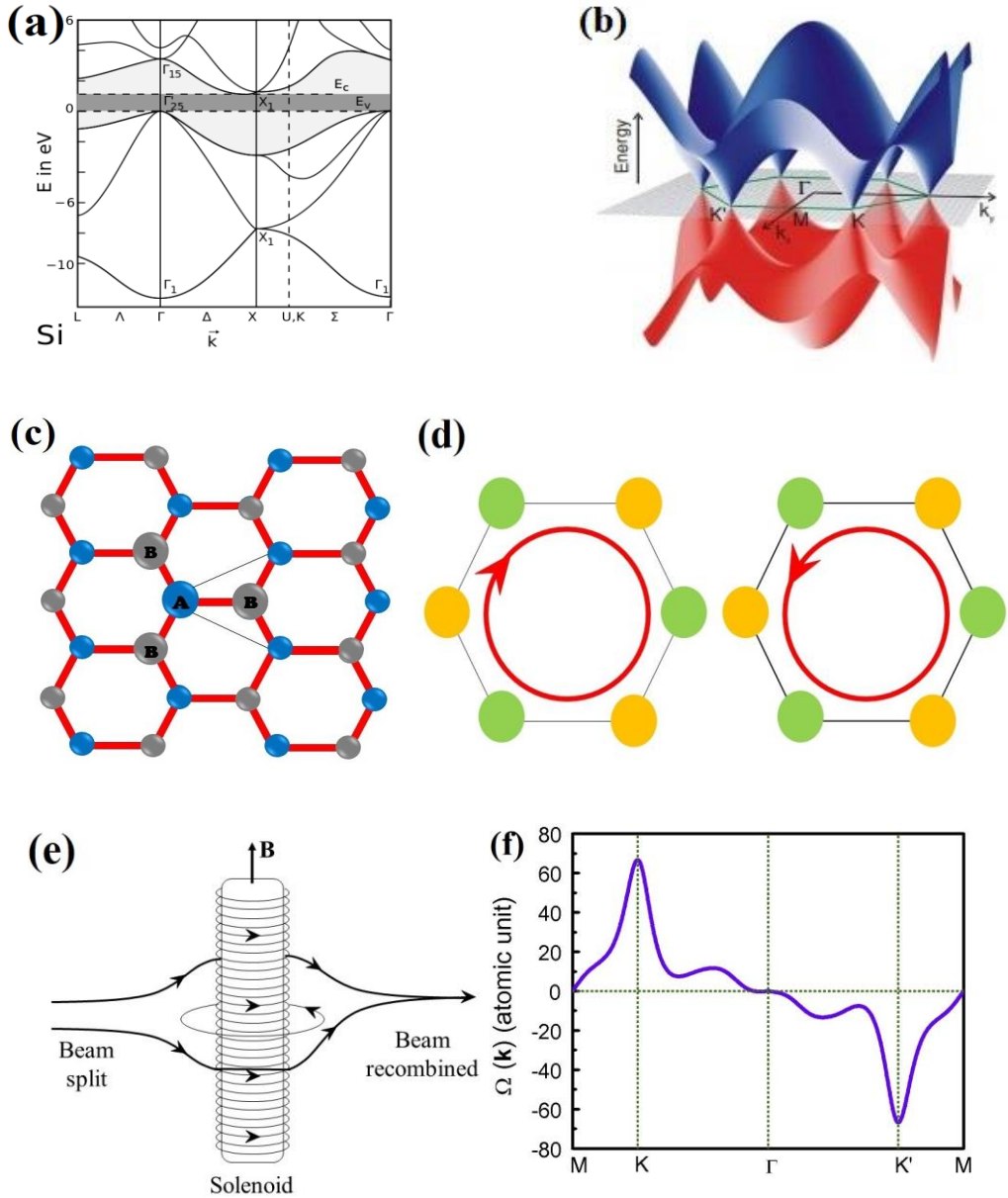

**Figure 1** (a) Energy band structure of silicon showing multiple valleys. (b) Schematic drawing of the characteristic Dirac cones of the graphene band structure.[25], (c) Hexagonal graphene lattice consisting of two triangular sublattices (A and B). (d) Self-rotation of the electronic wave packet in K and K′ valleys. (e) Pictorial view of Aharonov–Bohm experiment. (f) Barry curvature of $MoS_2$ along with high symmetry lines.[39]

## 2. Valley Physics

The valleys in crystalline solids could be equivalent or inequivalent depending on the nature of the unit cell. When there is some asymmetry in the unit cell valleys exhibit different characteristics. For example, in the hexagonal graphene lattice which can be considered as a superposition of two identical sub-lattices set off by the carbon-carbon bond length (fig 1c), two inequivalent (since the two sublattices are distinct) otherwise identical valleys are present at K and K′ points in the Brillouin zone. The K and K′ points in graphene are related to inversion and time-reversal symmetry.[40] Because of broken sublattice symmetry in 2D TMDCs the valleys located at K and K′ are degenerate but having opposite momentum.[20] To differentiate these valleys a new index, known as valley pseudospin ($\tau_z$), has been introduced.[26] This is called pseudospin because like spin it carries information, up/down or valley one or two. The values of ($\tau_z$) for TMDCs are ±1, which describe the position of particles: in K or K′ valley.[26] In 2D materials, there are two physical quantities- orbital angular momentum and Berry curvature that are different at K and K′ valleys.

### 2.1 Orbital magnetic moment (m)

The orbital magnetic moment is one of the valleys contrasting parameters in 2D materials.[41, 42] It can be regarded as the self-rotation of the electron wave packet clockwise in one valley (at K) and anticlockwise in the other valley (at K′) giving rise to an opposite value of m (Fig 1d). As the spin magnetic moments are different in the case of spin up and downstate, the same way, the orbital angular moment discriminates different valley states. The valley contrasting magnetic moment can be defined as[26]

$$m(k) = \tau_z \frac{3ea^2 \Delta t^2}{4\hbar(\Delta^2 + 3q^2 a^2 t^2)} \qquad (1)$$

The orbital magnetic moment is related to some fundamental symmetries (time-reversal symmetry and inversion symmetry) in a crystal. Time-reversal is the name of a particular symmetry transformation that stands for an inversion of the direction of time,[43] while spatial inversion symmetry refers to symmetry under a reversal of the direction of all the coordinate axes.[44] Both $m$ and $(\tau_z)$ are change sign under time-reversal symmetry operation. In contrary, only $(\tau_z)$ (not $m$) changes sign under spatial inversion. This implies that $m$ can only be zero when the system remains invariant under both time-reversal and inversion symmetry. The value of orbital magnetic moment can be nonzero only in systems with broken inversion symmetry. Although, the $K$ and $K'$ points are time reversed images of one another, orbital angular moment at these points in hexagonal 2D materials are inequivalent due to the violation of spatial inversion symmetry.

**2.2 Berry phase and curvature**

The adiabatic theorem in quantum mechanics states that a system will remain in its time-dependent ground state if the Hamiltonian is changed slowly. In reality, when the Hamiltonian of a system travers around a closed loop in the parameter space, an irreducible phase known as the Berry phase can be developed.[45] The Berry phase is the geometrical phase that acquired during the adiabatic evolution of a quantum state. The foremost features of the Berry phase are 1) it is geometrical, 2) it is gauge invariant, and 3) it has close analogies to gauge field theories and differential geometry.[46, 47] Berry phase effect is observed in crystalline solids having band structure.[48] Under the free electron approximation, the Hamiltonian for a single electron in a crystal lattice is expressed as

$$H = \frac{\hat{p}^2}{2m} + V(r) \tag{2}$$

where *V(r)* is the periodic potential. According to Bloch's theorem, the energy eigenstates of a periodic Hamiltonian can be written in the form:[49]

$$\Psi_{nk}(r+a) = e^{i.k.a}\Psi_{nk}(r) \qquad (3)$$

where *a* is the Bravais lattice vector, *n* is the band index, and $\hbar k$ is the crystal momentum. Although the Hamiltonian of the system is k-independent, the boundary condition depends on k. The Hamiltonian can make k-dependent through a unitary transformation of the form

$$H(k) = e^{-ik.r}He^{ik.r} = \frac{(\hat{p}+\hbar k)^2}{2m} + V(r) \qquad (4)$$

The part of the Bloch function is $u_n k(r) = e^{-ik.r}\psi_n k(r)$, which satisfies the periodic boundary condition

$$u_n k(r+a) = u_n(k) \qquad (5)$$

All such eigenfunctions belong to the same Hilbert space. Therefore, the Brillouin zone can be the parameter space of the transformed Hamiltonian $H(k)$ with $|u_n(k)\rangle$ as the basis function.

Since the k dependence of the basis function is inherent to the Bloch problem, various Berry phase effects are expected in crystals. For example, in crystals, if k is considered to be in the momentum space, then a Berry phase will be associated with the Bloch state:

$$\Phi_B = \oint_C dk.\langle u_n(k)|i\Delta k|u_n(k)\rangle = \oint_C dk.\Omega_n(k) \qquad (6)$$

The Berry curvature of the energy bands of crystalline solids is defined as

$$\Omega_n(k) = \Delta k \times \langle u_n(k)|i\Delta k|u_n(k)\rangle \qquad (7)$$

It depends on the wave function and hence an intrinsic property of the crystal. It plays a vital role in the accurate description of the dynamics of Bloch electrons.

In the Aharonov–Bohm effect that is a purely quantum mechanical phenomenon, an electromagnetic potential (φ, **A**) affects an electrically charged particle even if both the magnetic (B) and electric (Ɛ) fields are zero in that region.[50] After splitting into two, a beam of electrons is sent to each beam past the solenoid on a different side of it (Fig 1e). Suppose the solenoid contains a magnetic field (non-zero vector potential outside solenoid). In that case, the electronic state will acquire an additional phase (Φ), which is proportional to the magnetic flux ($\phi_m$) inside the solenoid.

$$\Phi = \frac{e}{\hbar} \oint A(r) = \frac{e}{\hbar} \phi_m \tag{8}$$

$$\phi_m = \int da.B \tag{9}$$

Where, **da** and **B** are the cross-sectional area of the solenoid and magnetic field, respectively. This effect is known as the magnetic Aharonov–Bohm effect. Making an analogy of Berry-phase effect with magnetic Aharonov–Bohm effect, Berry phase and Berry curvature can be considered as magnetic flux and magnetic field in the momentum space, respectively.[51]

The value of the Berry curvature is nonzero for crystals with broken time-reversal or inversion symmetry. 2D TMDCs have non-zero Berry curvature because of the absence of an inversion center in the structure. In these materials, maximum Berry curvatures are observed at *K* and *K′* points (Fig 1f).[39] Another example of a system where both symmetries are not simultaneously present is monolayer graphene with staggered sublattice potential, where inversion symmetry is broken.[52] The Berry curvature of this system possesses opposite signs at valley $K_1$ and $K_2$ due to time-reversal symmetry.[26, 53] While the energy gap approaches zero, the Berry phase of an electron after completing one circle around the valley becomes ±π. The intrinsic graphene sheet exhibits a Berry phase of π.[4, 29]

**2.3 Valley selection rule**

When a Bloch electron travels adiabatically in a non-degenerate energy band, in general, the real-space dimension of the associated wave packet is much larger than the lattice constant but much smaller than the length scale of the external perturbation.[54, 55] Therefore, the wave packet and the wavevector of the electron can be considered independently. In that case, we can describe the motion of electrons in the crystal lattice by the semi-classical equation where the wave function is a Bloch function, and the mean velocity is related to the gradient of electronic energy of the band.[56, 57] When the periodicity, as well as the applied magnetic field, is taken into consideration, there exists an anomalous velocity due to the Berry curvature of the electronic band. The Berry curvature of the Bloch states exists in the absence of the external fields and manifest in the quasi-particle velocity when the crystal momentum is moved by external forces.[58, 59]

The semi classical dynamics of a Bloch electron under applied electric and magnetic field can be expressed as

$$\frac{dr}{dt} = \frac{1}{\hbar}\frac{\partial E_n(k)}{\partial k} - \frac{dk}{dt} \times \Omega_n(k) \tag{10}$$

$$\hbar\frac{dk}{dt} = -e\varepsilon - e\frac{dk}{dt} \times \vec{B} \tag{11}$$

where $E_n(\boldsymbol{k})$ The energy dispersion of the n$^{th}$ band, $\boldsymbol{r}$ is the crystal momentum and position of the electron wave packet, and $\mathcal{E}$ and $\vec{B}$ represent the applied electric and magnetic field, respectively. The Berry curvature can be defined in terms of Berry connection

$$A_n(k) = \langle u_n(k) | i \nabla k | u_n(k) \rangle \tag{12}$$

as

$$\Omega_n(k) = \nabla_k \times A_n(k) \tag{13}$$

According to Kubbo formula,[56, 60, 61] the Berry curvature of the Bloch states can also be written as

$$\Omega_n(k) = i \frac{\hbar^2}{m^2} \sum_{n' \neq m} \frac{\langle u_n | v | u_{n'} \rangle \times \langle u_{n'} | v | u_n \rangle}{\left[ E_n^0 - E_{n'}^0 \right]^2} \tag{14}$$

Where, $E_n^0$ is the energy dispersion of the n$^{th}$ band and $\mathbf{v}$ is the velocity operator. Because the equation of motion remains invariant under symmetry operations, $\Omega_n(k) = -\Omega_n(-k)$ under time-reversal symmetry, and $\Omega_n(k) = \Omega_n(-k)$ under inversion symmetry. Therefore, valley-contrasting properties appear when one of the symmetries is broken.

The Berry curvature gives rise to an anomalous transverse velocity (Hall velocity) in a 2D crystal in the presence of an electric field:[48]

$$v_\perp = \frac{dk}{dt} \times \Omega_n(k) \tag{15}$$

As the direction of electrons velocity in opposite valleys is opposite, the motion of electrons of a particular valley can be selected by changing the direction of the applied electric field.

The electron energy dispersion can be written in terms of orbital angular momentum as

$$E_n(k) = E_n^0(k) - m_n(k).B \qquad (16)$$

Where,

$$m(k) = i\frac{e\hbar}{m^2}\sum_{n'\neq n}\frac{\langle un|v|un'\rangle \times \langle un'|v|un\rangle}{\left[E_n^0 - E_{n'}^0\right]} \qquad (17)$$

The valley magnetic moment of a valley carrier interacts with an external magnetic field giving rise to a valley Zeeman effect similar to the spin Zeeman effect. Therefore, an external magnetic field could be an additional way of controlling VDF. Nonetheless, due to the finite orbital magnetic moment valley carriers interact differently with light of left and right circularly polarized light.[35, 36, 62] Such optical circular dichroism due to the orbital magnetic moment provides optical selection rules for valley carriers.[20]

In monolayer TMDCs, contrasting $\boldsymbol{\Omega}$ and $\boldsymbol{m}$ values in the $\pm K$ valley gives rise to valley selection rules. In the tight-binding approximation, the Hamiltonian of a single TMDC sheet depends on nearest-neighbor hopping energy (hopping integral) t and bandgap energy $\Delta$. The low-energy description of the Hamiltonian near the Dirac points is given by[37, 63]

$$H = at(\tau_z k_x \sigma_x + k_y \sigma_y) + \frac{\Delta}{2}\sigma_z \qquad (18)$$

Where, $\sigma$ is the Pauli matrix accounting for the sublattice index and the lattice constant. According to this model, the valley-contrasting Berry curvature takes the form

$$\Omega_c(k) = -\hat{z}\frac{2a^2t^2\Delta}{(2a^2t^2k^2+\Delta^2)^{3/2}} \quad (19)$$

As the values of $\tau_z = \pm 1$, $\Omega_c$ assumes equal but opposite values in $K$ or $K'$ valleys of monolayer TMDCs. The valley contrasting $\Omega_c$ gives rise to a current of the carriers with the sign depending on the valley index ($\tau_z$) while exposed to an in-plane electric field known as the valley Hall effect. The photoexcited electrons and holes move to the two opposite edges of the TMDC sheet developing a transverse bias (Fig 2a).[37, 64] This phenomenon is known as velley Hall effect.

According to massive Dirac fermion model, the orbital angular moment can be expressed as

$$m(k) = -\hat{z}\frac{2a^2t^2\Delta}{4a^2t^2k^2+\Delta^2}\frac{e}{2\hbar}\tau_z \quad (20)$$

In the low-energy limit (k→0), the orbital magnetic moment is[26]

$$m(m_{1,2}) = \tau_z \mu_B^* \qquad \mu_B^* = \frac{e\hbar}{2m_e^*} \quad (21)$$

Where, $m_e^* = (2\Delta\hbar^2)/(3a^2t^2)$ is the effective mass of the electron at the bottom of the band. Therefore, monolayer TMDCs exhibit valley contrasting magnetic moment which can be detected through an applied magnetic field (Fig 2a).[37]

The optical circular dichroism in 2D TMDCs is given by

$$\eta(k) = -\frac{m(k).\hat{z}}{\mu_B^*(k)} = -\frac{\Omega(k).\hat{z}}{\mu_B^*(k)}\frac{e}{2\hbar}\Delta(k) \quad (22)$$

In the Brillouin zone at high symmetry points, the Bloch states are invariant under a q-fold discrete rotation about the direction of light propagation: $\Re\left(\frac{2\pi}{q},\hat{z}\right)|\Psi_{c(k),k}\rangle = e^{-i\frac{2\pi}{q}l_{c(v)}}|\Psi_{c(v),k}\rangle$. The azimuthal selection rule for interband transitions by circularly polarized light is $l_v \pm 1 = l_c + qN$ [28] where, $l_v$ and $l_c$ are azimuthal quantum numbers. In TMDCs, the right and left-circularly polarized lights couple to interband transitions at K and K′ valleys, respectively.[54]

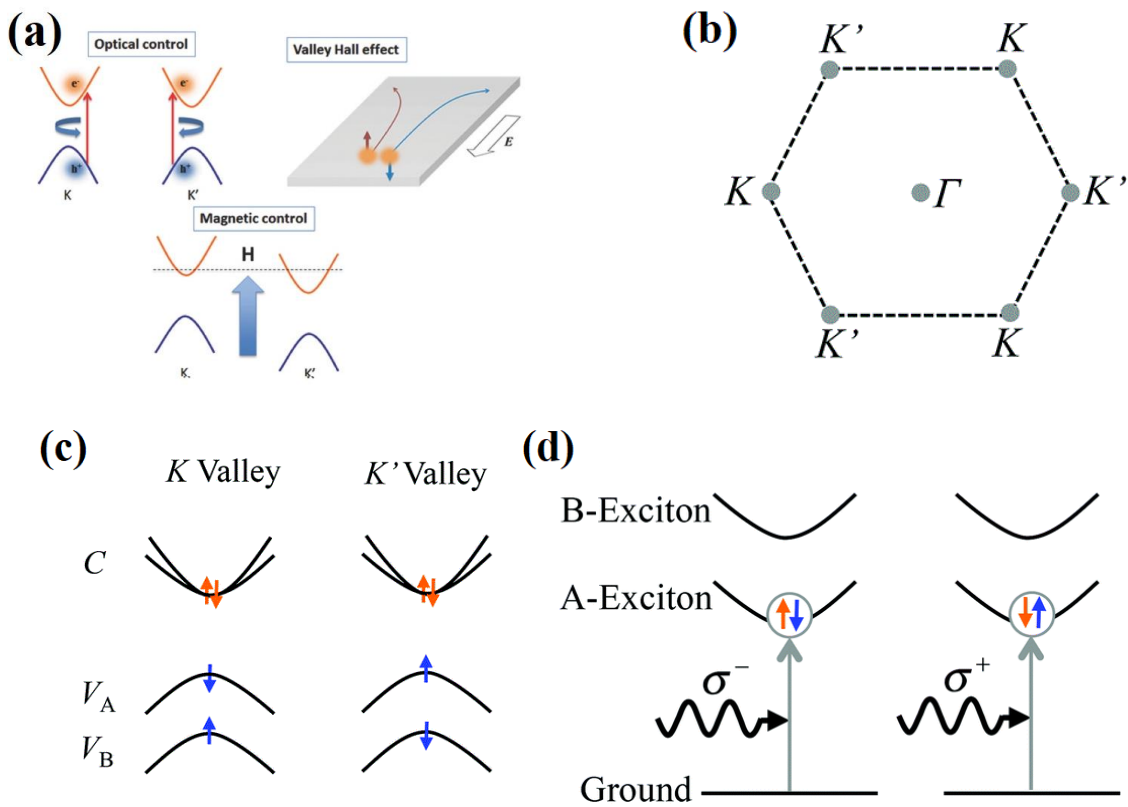

**Figure 2** (a) Various methods (i.e., optical, magnetic, and electrostatic) to control the VDF.[37] (b) K and K' valleys in momentum space.[65] (c) Electronic band structures at K and K' valleys with their spin orientations (up and down arrows).[65] (d) Spin orientations of A excitons (at K and K' valleys) and their coupling with circularly polarised light.[65]

The lattice of 2D TMDC crystals has no inversion symmetry and therefore it's six energy valleys in momentum space constitute two inequivalent sets (K and K′) (Fig 2b). The strong spin-orbital coupling in d orbitals of the chalcogenide atoms introduces large splitting in the valance bands of such crystals.[66, 67] The spin splitting at different valleys is opposite due to time-reversal symmetry and the lack of inversion symmetry in space. The spin of holes in the $V_A$ band are −3/2 and +3/2, respectively, while the spin in the $V_B$ band assumes the values +1/2 and -1/2 in the respective valleys (Fig 2c).[65] Because of this unique spin feature, the spin and valley degrees of freedom of holes are inherently coupled in the valance bands.[20, 68] However, the degeneracy in the conduction bands near the band minima in TMDCs (Fig 2c)[65] provides interesting spin and valley structures, which gives rise to the spin- and valley-selective optical coupling.[69] The holes in the $V_A$ and $V_B$ bands are combined with electrons in the conduction band and form A and B excitons, respectively (Fig 2d).[65] In the K valley, a spin-down hole (−3/2) and either a spin-up (+1/2) or a spin-down (−1/2) electron constitute A-exciton. The spin of the exciton in the first case is -1(down) and hence couples to right circularly polarized photons (Fig 2d). The second exciton is a dark exciton with spin -2 and does not couple to photons. On contrary, the bright A-exciton in the K′ valley is formed out of a spin-up hole (+3/2) and a spin-down electron (−1/2) having a spin of +1 (spin-up exciton), and hence couples to left circularly polarized photons (Fig 2d). Hence, both spin- and valley-polarized excitons can be created in 2D TMDC crystals using circularly polarised light of different kind.[36, 69]

## 3. Valley carriers

Like charge and spin, valley information can also be transported through materials. There are many carriers of valley pseudospin, for example, free electrons and holes, doped electrons and holes, quasi-particles like neutral, charged, and bound excitons. Because of the availability of various valley information carriers, several different ways can be used to control transport of valley polarization.

**3.1 Electrons and holes**

Hall currents flow transversely to the applied electric field in graphene placed on top of hexagonal boron nitride (h-BN) even in the absence of a magnetic field.[70] The opposite Hall currents arise due to carriers located in opposite valleys.[20, 26] The inversion symmetry broken by a superlattice potential induced by h-BN substrate gives rise to the Hall currents.[71, 72] Pure valley current flows through dual-gated bilayer graphene (BG) and produces a large non-local resistance that scales cubically with the local resistivity (Fig 3).[73] A pronounced non-local response in the resistance has also been observed as a result of the topological transport of the valley pseudospin in bilayer graphene (BG) (Figure 3d-g).[31] In BG, carriers that are doped by applying an out-of-plane electric field carry the valley information.[31, 73] These carriers of valley pseudospin can easily be injected using an electric field and controlled by electric or magnetic fields. The lifetime of these valley carriers is very long (~ns).[74]

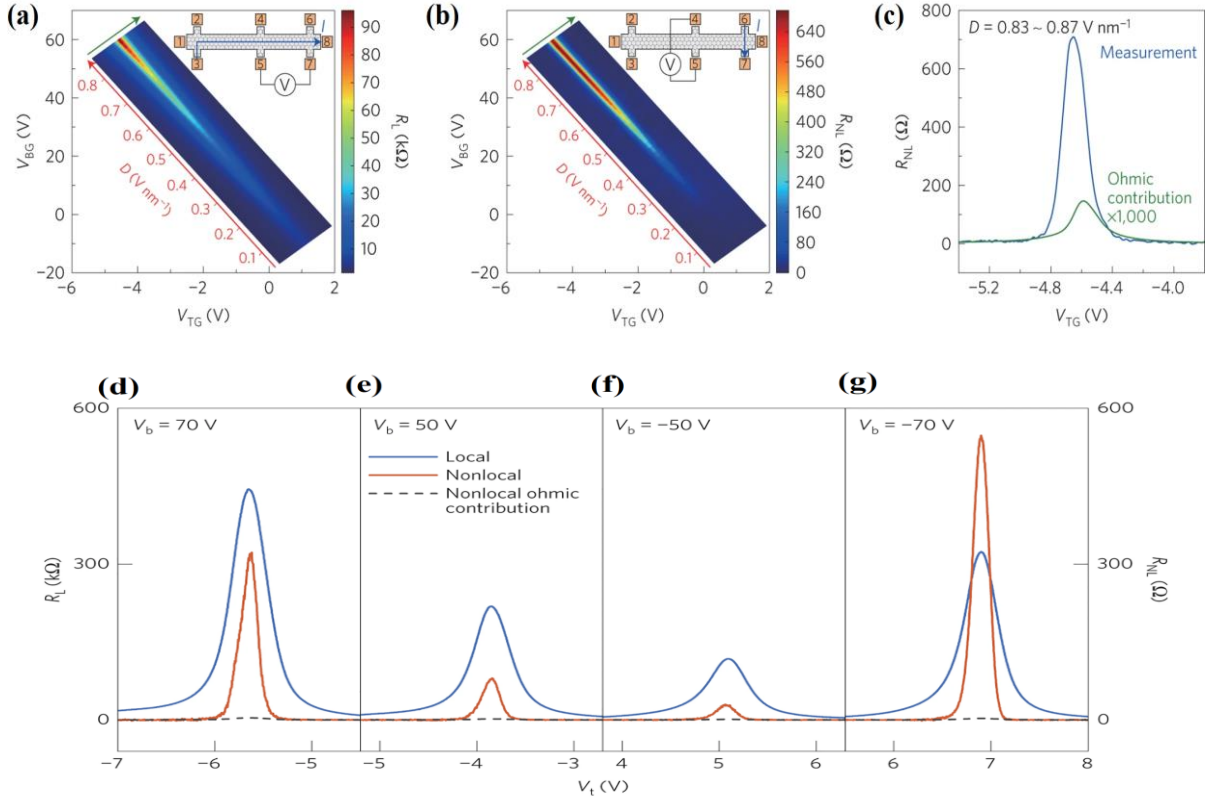

**Figure 3** (a), (b) Measured local ($R_L$) and non-local ($R_{NL}$) resistances. Dependence of $R_L$ and $R_{NL}$ on gate voltage, respectively. Back and top gate displacements are defined as, $D_{BG} = \varepsilon_{BG}(V_{BG} - V_{BG}{}^0)/d_{BG}$ and $D_{TG} = -\varepsilon_{TG}(V_{TG} - V_{TG}{}^0)/d_{TG}$ respectively.[73] (c) Comparison of experimentally measured $R_{NL}$ with a calculation of ohmic contribution.[73] (d-g) Variation of $R_L$ and $R_{NL}$ with $V_t$ (voltages applied on top gates) at fixed $V_b$ (voltages on bottom gates) at varying voltages. The sample dimensions were 5 by 1.5 μm. The dashed lines represent the expected ohmic non-local contribution.[31]

Optically injected electron and hole exhibit valley polarization in TMDs. In monolayer TMDs, the electrons (injected through optical excitation following the valley optical selection rules) from different valleys are moved in opposite directions across the sample and produce valley current.[64, 75] Optically generated carriers are also responsible for the valley polarization in bilayer TMDs.[76-78] The advantages of such carriers are they can easily be created using light and have a long lifetime.[79-81] The valley current can be generated using an electric field[82] and controlled by electric and magnetic fields.

**3.2 Intralayer excitons**

In the low dimension material, due to the weak electrostatic screening and large coulomb interactions quasi-states (excitons, trions, and biexcitons) are formed.[83-85] An exciton is a combined state of an electron in the conduction band and a hole in the valence band.[86-89] In the doped semiconductor, the neutral exciton can be combined with an extra electron or hole to form a charged exciton or trion.[90, 91] The enhanced Coulomb interaction in 2D materials (e.g., TMDCs) leads to the formation of biexciton.[92, 93]

Monolayer TMDCs possess a direct bandgap and exhibit a great resonance in optical response like absorption and photoluminescence (PL) due to the formation of neutral ($X^0$) and charged ($X^{+/-}$) excitons (Figure 4a). The typical absorption spectra of monolayer TMDCs consist of several characteristic peaks due to exciton resonance and interband transitions.[94] The theoretical studies for such systems have reported unusual features in the density of states, i.e., 'band nesting' (Fig 4b).[95] According to these studies, the location of band nesting region is near the midway between Γ and K′, where the conduction band and the valence band are parallel to each other. The conduction band minima (CBM) and valence band maxima (VBM) are both located at the K/K′ point of the Brillouin zone. Hence, the optical response of monolayer TMDCs is dominated by valley excitons (bound electron-hole pairs in the valleys).

Figure 4c shows the PL and differential reflectance spectra of monolayer $MoS_2$, $MoSe_2$, $WS_2$, and $WSe_2$ on a quartz substrate.[96] The peaks A and B are associated with the excitons that reside at K and K′ points. A strong absorption peak is observed due to the point where the conduction and the valence bands are nested, i.e., C peak in Fig 4c. As a consequence of the strong quantum confinement, the exciton binding energy in TMDCs is exceptionally high, about 0.5 eV.[94, 97] These excitons are bright excitons in which electrons and holes remain in the same valley. Therefore, like free electrons and holes, bright exciton has a binary valley pseudospin. In excitons, the Coulomb interactions between charges are of two types: direct and exchange. While the binding energy of excitons is predominantly due to the direct part, the exchange Coulomb interaction results in diagonal energy shift and off-diagonal coupling on the valley configurations.[98] The electron-hole exchange interaction between the excitons in two valleys is equivalent to an in-plane effective magnetic field on the valley pseudospin.[54, 99] This effective magnetic field gives rise to the valley–orbit coupling and the bright exciton dispersion is split into two branches having in-plane valley pseudospins (Figure 4d).[98]

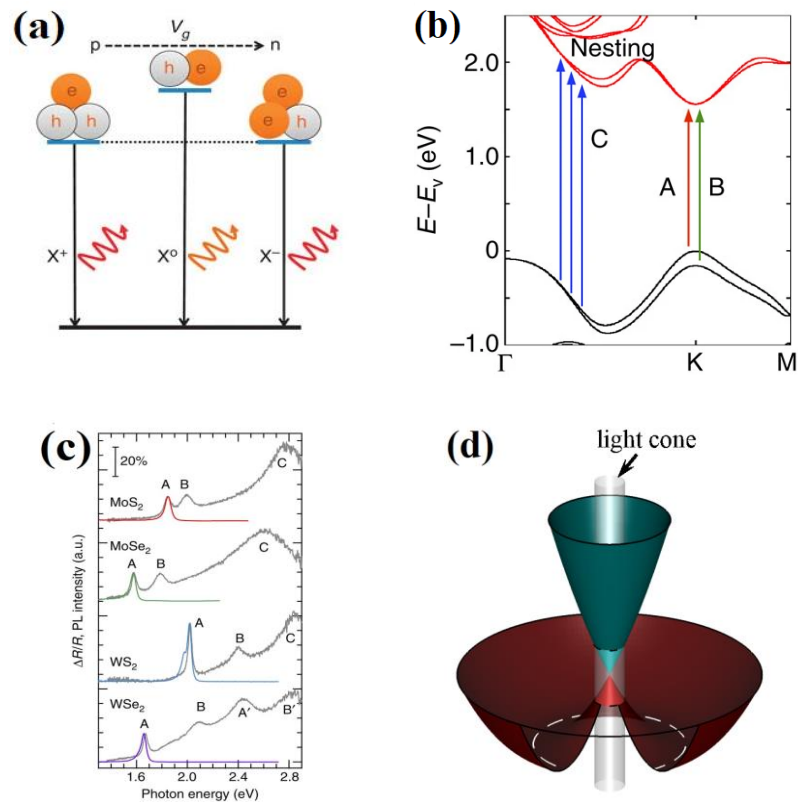

**Figure 4** (a) Illustration of exciton and trion states in TMDCs and associated transitions.[100] (b) The calculated energy band diagram of monolayer $MoS_2$. The transition in A, B, and the band nesting are represented by arrows.[96] (c) PL spectra (red, green, blue, and purple curves) and differential reflectance spectra (grey curves) of $MoS_2$, $MoSe_2$, $WS_2$, and $WSe_2$.[96] (d) Valley–orbit splitting of bright exciton by the intervalley electron-hole exchange interaction.[99]

Because of the substantial resonant excitonic absorption[101] of monolayer TMDCs, valley excitons can play an important role in TMDC based valleytronics devices.[98] These devices are expected to be easily initialize using light[34, 102] and readout can be done by photon polarization.[35, 102] Valley excitons can be controlled by both magnetic[103-105] and optical fields.[106, 107] The major issue with this kind of valley carrier is short (few nanoseconds) exciton lifetime[108, 109] and very short (few picoseconds) valley lifetimes.[110, 111]

Nonetheless, recent theoretical and experimental reports suggest that dark excitons can play an important role in determining the valley degree of polarization in 2D materials.[97, 112, 113] Such exciton states are a huge reservoir for valley polarization and intra valley scattering between bright and dark excitons. This helps maintain a Boltzmann distribution of the bright exciton states in the same valley.[112] In TMDCs, the degree of circular polarization is directly related to the alignment of bright and dark exciton states. The PL quantum yield of tungsten-based TMDCs was found to be increased at higher temperatures. This observation can be explained by the presence of dark exciton states having energy lower than the bright state.[114]

Dark excitons can be classified as spin-forbidden and momentum-forbidden dark excitons (Figure 5a ). The spin-orbit interaction removes the spin degeneracy of the conduction band of TMDCs leading to different spins for conduction band electrons. The signs of spin for conduction band electrons are different for tungsten- and molybdenum-based TMDCs,[115] as a result, spin-forbidden dark excitons are formed in such systems (Figure 5a). The energy of the spin-forbidden dark exciton is less than the bright exciton in molybdenum-based TMDCs, whereas its energy is lower than the bright exciton in tungsten-based TMDCs. These dark states can be brightened by applying strong in-plane magnetic fields, which mix the spin-split bands through the Zeeman effect.[114, 116] The result is the relaxation of the spin-selection rule and brightening of dark excitons. The PL from dark excitons has been demonstrated in low-temperature magneto-PL experiments. In the presence of a magnetic field, the PL spectra of $WS_2$ monolayers exhibit a doublet structure because of the splitting of the spin-forbidden states.[114] Moreover, the spin-forbidden dark excitons in $WSe_2$ have been activated on silver surfaces through coupling to surface plasmon modes.[117]

The excitons that are formed due to Coulomb interaction between electrons and holes located at different valleys are momentum-forbidden dark excitons. In tungsten-based TMDCs, the dark excitons are formed with the electron at the $\Lambda$ valley and the hole at the K (Figure 5a).[118] These excitons can emit PL through phonon-assisted radiative recombination.[119] Another way of getting PL from these dark excitons could be exciton-molecule coupling.[120] In the presence of molecules of high-dipole moment, translational symmetry of the TMDCs gets disturbed making the K→Λ transition allowed by supplying the required momentum. Theoretical studies suggest that dark excitons can be created in the monolayer of TMDCs ternary alloys ($Mo_{1-x}W_xX_2$).[113, 121]

**3.3 Interlayer excitons**

The requirement of a long valley polarization lifetime for valleytronics device applications could be fulfilled by creating interlayer excitons. In vertically stacked 2D heterostructures even when electrons and holes are located in different TMDC layers, interlayer excitons are formed due to the strong Coulomb interaction (Figure 5b).[122-128] A type-II band alignment can be seen in a certain combination of TMDCs while vertically stacked.[128-132] In such vertical heterostructures, interlayer charge transfer occurs because of the staggered band alignment and large band offsets.[122, 123, 133] As a result of the absence of depletion region in such p-n junctions, electrons and holes are largely localized to different layers.[134-136] These excitons are stable at room temperature and do not dissociate even in the presence of external fields due to high binding energies (~100 meV).[128, 137, 138] Interlayer excitons have been identified in several TMDC heterostructures through their PL signature.[123-125, 139, 140] The mechanical stacking of one 2D layer on top of another causes the mismatch in the lattice constants, and a twist between the two layers of a heterostructure leads to Moiré pattern in real space (Fig 5c).[141, 142] The pattern depends on the twist angle between two layers; therefore, the optical response of the heterostructures can be controlled by tuning the relative orientation of the layers.[143-145]

The conduction and valence bands of the two layers of a heterostructure are different, which makes interlayer excitons momentum-space indirect. These excitons are also spatially indirect as constituent electrons and holes reside on different 2D layers. The result is a much longer lifetime of the interlayer exciton than the intralayer exciton in the monolayer.[123, 124] The interlayer exciton lifetime in $WSe_2$-$MoSe_2$ heterostructures is an order of magnitude longer than intralayer excitons.[123, 124] A valley lifetime of 40 nanoseconds has been achieved by creating interlayer exciton spin-valley polarization using circularly polarized light in $WSe_2$-$MoSe_2$ heterostructures.[146] The lifetime of interlayer excitons can be increased further by making a strongly confining trap array using a $WSe_2/WS_2$ heterostructure on a patterned substrate.[147] Interlayer excitons can be generated and controlled by applying an electric field. One of the most prominent examples of this is an optical and electrical generation of long-lived neutral and charged interlayer excitons in hexagonal boron nitride (h-BN)–encapsulated vdW heterostructures of $MoSe_2$ and $WSe_2$.[139, 148]

Two kinds of interlayer excitons (direct and indirect) have been observed in van der Waals heterostructures.[140] These excitons exhibit PL lifetimes of several tens of nanoseconds. In heterostructures of monolayers of $MoSe_2$ and $WSe_2$, the low-energy interlayer exciton is indirect both in real and reciprocal space. In contrast, the high-energy interlayer exciton is only indirect in real space.[140] The interlayer dark excitons in 2D heterostructures can survive for a very long time (~microsecond).[149] Such long-lived excitons are produced by the magnetic field suppressed valley mixing, and hence can be controlled by the magnetic field.

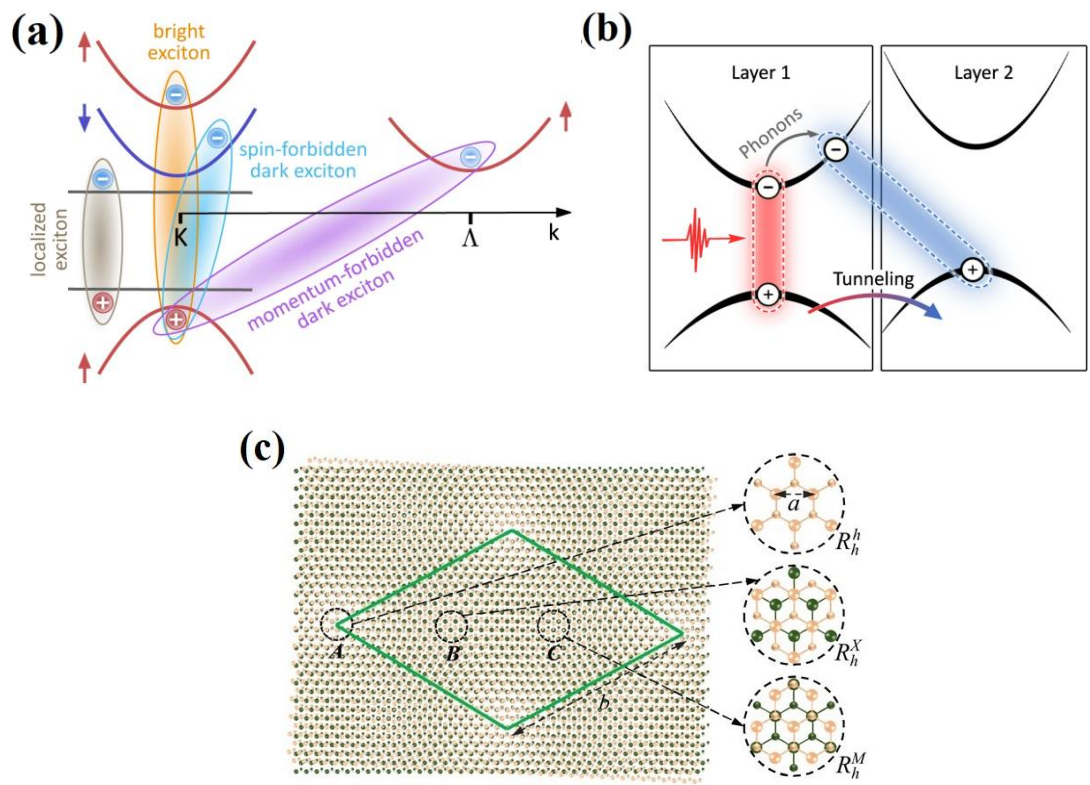

**Figure 5** (a) Schematic representation of bright exciton, spin forbidden dark exciton, and momentum forbidden dark exciton in atomically thin nanomaterials.[97] (b) Interlayer excitons are formed by the electrons and holes present in different layers.[97] (c) Moiré pattern in an heterostructure ($MoX_2$/$WX_2$). A supercell is indicated by green quadrilateral. Insets are close-ups of three locals.[141]

**3.4 Bound excitons**

Bound excitons are formed in 2D TMDCs when electrons and holes are trapped in defects that are formed due to impurities and/or strain in samples. The defects can affect the momentum-space valley properties by localizing excitons in real space and hence valley polarization can be controlled by manipulating defects in 2D vdW materials. PL from bound excitons is observed at low temperatures.[150] These emission lines are spectrally narrow and are observed below the broad PL peak corresponding to the bright excitons.[151-153] In fact, the strong single-photon emission from TMDC monolayers appears due to the existence of bound excitons.[154, 155] The narrow emission lines emitted from both CVD-grown and exfoliated samples exhibit a fine structure doublet.[151, 152, 155] In the absence of magnetic field, the doublet is cross-linearly polarized.[152]

A large Zeeman splitting (much more extensive than that for intrinsic excitons) is observed in $WSe_2$ in the presence of a transverse magnetic field (applied in the Faraday geometry).[152] The increase of spectral splitting in the doublet with the strength of the applied magnetic field suggests shifting the selection rule from linearly polarized to circularly polarized.[151, 155, 156] At a high magnetic field (> 5 T) the selection rule becomes similar to that of recombination of excitons in the regular $WSe_2$ lattice. In the absence of magnetic field, there is a mixing of the circularly polarized transitions at two valleys (K and K′) which results in two linearly polarized transitions with an energy splitting. The linearly polarized doublet basically originates due to the asymmetry of the confining potential.[152] Because of the dominant Zeeman splitting over the exchange energy, circularly polarized selection rules become relevant at high magnetic fields. This means bound exciton states are probably the superposition of the intrinsic valley exciton states, and hence single localized valley exciton can be exploited as a basis for valleytronics. Although bound excitons exhibit a long lifetime (~ ns)[152, 154, 155] and can be controlled by magnetic[152, 155] and optical[152, 157] fields, it could be a real challenge to manipulate valley properties by tuning defects.

**3.5 Trions**

The charged excitons (trions), which are three-body bound states of electrons and holes, could potentially function as carriers in valleytronics[158-160] due to their much longer valley lifetime than excitons [158, 160], robust valley polarization, and coupling to the additional charge.[161] There are two types of trions: bright and dark.[162] The bright trions can have the intravalley singlet state where carriers are located within one valley and the intervalley triplet state with an exciton in one valley and a charge carrier (electron/hole) in another valley. The fine structure of trions appears due to exchange interaction which splits energy levels of singlet and triplet trions.[163] The energy of the singlet trion state is lower than that of the triplet state.[100, 164-166] In doped semiconductors, after the recombination of electrons and hole of negatively charged trions the background electron gas becomes spin-polarized.[167-169] In TMDCs, the residual charge can exhibit definitive valley index and spin orientation and therefore, trions can be exploited to manipulate the VDF.[170, 171] The spin and valley lifetimes of trions are generally long because the intervalley scattering of the extra charge present in trions needs simultaneous transfer of a large momentum and spin-flip. Trions in 2D TMDCs exhibit a relaxation time of tens of picoseconds, which is much longer than ultrafast exciton relaxation time.[110, 172, 173] The chemical doping of monolayer $MoS_2$ leads to a large increase of the trion nonradiative lifetime.[174] A recent study demonstrates that chemical doping causes the conversion of excitons into trions in $WS_2$ monolayers, where the emission becomes dominated by trions with a strong valley polarization.[175] As a result of the chemical doping, exciton emission is strongly quenched but highly valley-polarized.

The excitation through a linearly polarized light creates a coherent superposition state of two trion valley conFigurations.[176] In the first configuration of trion, there exists a $\sigma^+$ photon and spin-down electron, whereas the second conFiguration consists of a $\sigma^-$ photon and spin-up electron. The superposition of these trion states leads to a spin-photon entangled state, $|\sigma^+\rangle|\downarrow\rangle + |\sigma^-\rangle|\uparrow\rangle$. The trion valley polarization cannot be detected through linearly polarized emission due to the orthogonal spin states. The trion valley coherence has been resonantly generated and detected in monolayer MoSe$_2$ using 2D coherent spectroscopy (three-pulse four-wave mixing technique).[176, 177] When an in-plane electric field is applied, an anomalous transverse motion is induced which gives rise to the Hall effects of trion.[98, 178] The strong polarization of the trion complexes in a WS$_2$ monolayer encapsulated in hBN under an applied magnetic field suggests that magnetic fields could be used to control the trion valley polarization.[179] The valley polarization in 2D TMDCs can be manipulated by modulating the electron-hole exchange interaction experienced by the exciton and trion by applying an electric field.[180]

## 4. Simple techniques to measure valley phenomenon

### 4.1 Polarisation resolved photoluminescence (PL)

The valley property of 2D materials can be investigated by monitoring the contrast between valley pseudospins at K and K′ valleys. Because the two valleys absorb left- and right-circularly polarized light differently, the basic mechanism generally used in this context is circularly polarized optical excitation.[26, 28] The valley polarization of the materials showing PL is measured via the circular dichroism of their PL.[181] In this approach, the polarization-sensitive PL measurements are carried out following the selective excitation of electrons in one of the valleys by circularly polarized light. The main aim here is to assess, the k-resolved degree of optical polarization or the degree of circular polarization, which can be defined as

$$\rho = \frac{I(\sigma^+) - I(\sigma^-)}{I(\sigma^+) + I(\sigma^-)} \quad (23)$$

Where, $I(\sigma^+)$ and $I(\sigma^-)$ are PL intensity of the right- and left-handed circular components.

The intensities of the PL with σ⁺ (left-handed) and σ⁻ (right-handed) polarization of the monolayer MoS$_2$ at low temperature (83 K) are much different (Fig 6a). Upon excitation by a σ⁺ light at 1.96 eV, the monolayer MoS$_2$ exhibits a circular polarization of ~50%.[36] However, reported experimental values of polarization are not consistent.[34-36, 182] The values of $\rho$ vary from as high as 100% to 32%.[34, 35, 182] The helicity parameter ($\rho$) is intimately linked to the A-exciton PL energy. It was found to be 100% for PL energies in the range 1.90–1.95 eV and decreases to 5% below 1.8 eV (Fig 6b).[35] Nonetheless, the degree of circular polarization of MoS$_2$ decreases as a function of photo-excitation energy and possesses a very high value for resonant excitation near the bandgap. In fact, it decreases following a powerlaw as the excitation energy increases (Fig 6c).[183] Kioseoglou et al.[183] proposed a phonon-assisted intervalley scattering-based model to explains the variation in reported values of the degree of circular polarization for MoS$_2$.

The valley polarization of the monolayer WS$_2$ has been explored through polarization-dependent PL.[78, 184] At the near-resonance condition, the monolayer exhibits a degree of circular polarization of 40% at 10 K (Fig 6d).[184] It is insensitive to PL energy but decays with increasing temperature and drops to 10% at room temperature. However, it decreases as the excitation energy changes from the near-resonance to the off-resonance. Interestingly, robust PL circular dichroism was observed for bilayer WS$_2$ under both resonance and off-resonance excitations.[184] The degree of circular polarization of bi-layer WS$_2$ under near-resonant excitation is 95% at 10 K and decreases to 60% at room temperature. The robustness of the valley polarization in bilayer WS$_2$ is the result of the coupling of spin, layer, and valley degrees of freedom. According to Nayak et al.[78], the effects of shorter exciton lifetime, smaller exciton-binding energy, extra spin-conserving channels, local spin polarization, and selective valley excitation can also contribute to the robustness of circular polarization in bilayers. Moreover, the PL measurements identified high valley polarization in other monolayer TMDs.[185, 186]

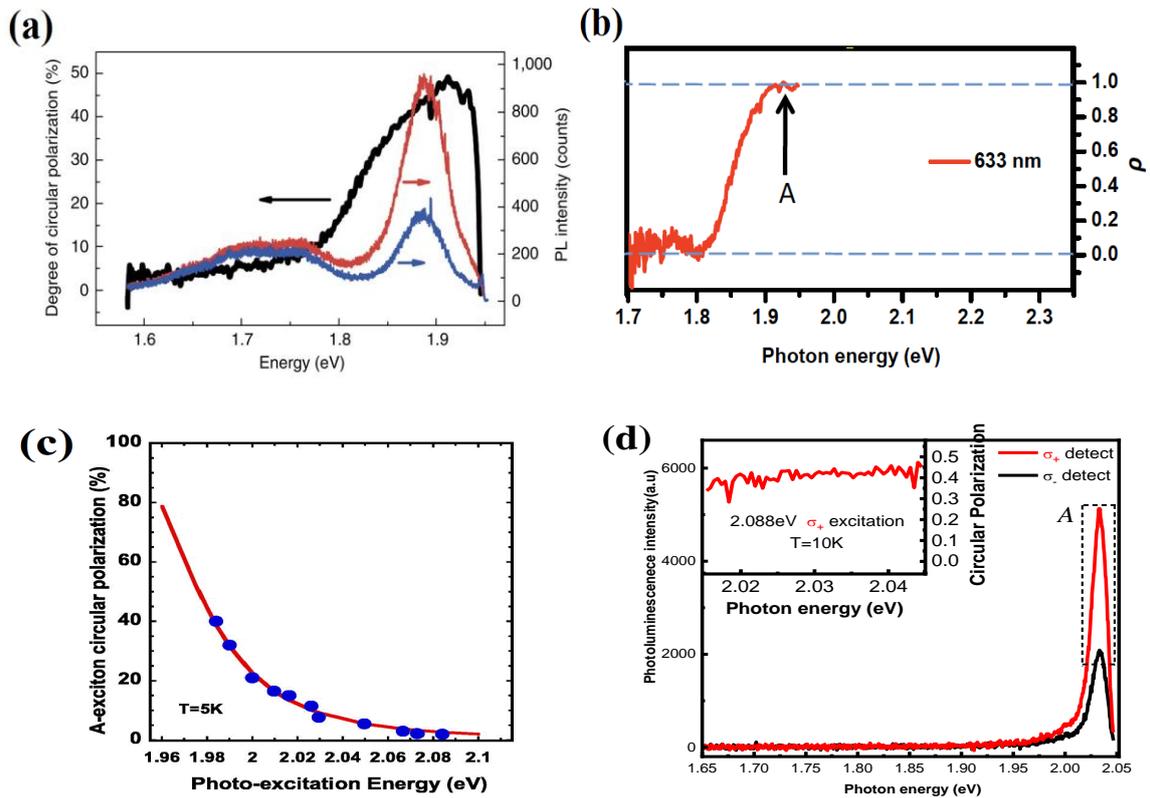

**Figure 6** (a) Circularly polarised PL spectra with σ+ (left-handed) (red) and σ− (right-handed) (blue) polarization of monolayer $MoS_2$ at 83 K along with the degree of circular polarisation (black).[36] (b) The helicity parameter (ρ) corresponding to the A-exciton of monolayer $MoS_2$ at an excitation photon energy of 1.96 eV.[35] (c) Degree of circular polarization of A-exciton of a single layer of $MoS_2$ as a function of the photo-excitation energy. A solid line is the fitting with a power-law. Inset shows dipole allowed optical transitions.[183] (d) Polarization resolved PL spectra with σ+ (red) and σ- detection (black) under near-resonant σ+ excitation (2.088 eV). The degree of the circular polarization at the PL peak is presented in the inset.[184]

## 4.2 Valley hall effect

As discussed in the previous sections, there could be a valley-dependent Berry phase effect that leads to a valley contrasting Hall transport. The charge carrier in the K and K′ valleys turn into opposite directions perpendicular to an in-plane electric field giving rise to the valley Hall Effect. This effect can be used to identify valley polarization in 2D materials.[31] In the case of a net valley polarization, a transverse Hall current will be generated upon the application of an in-plane electric field in a valley filter device.[17] This Hall current will cause a measurable transverse voltage which can give a local mapping of the valley- polarization of the sample (Fig 7a).[26] The valley-Hall effect (VHE) can create an electrical response in remote regions, which can be exploited to experimentally identify the presence of Hall currents.[70]

Graphene has Berry curvature due to the presence of a nonzero Berry's phase.[187] When the crystallographic axes of graphene and hexagonal boron nitride (hBN) are aligned, the inversion symmetry has broken and a finite Berry curvature is developed.[72, 188] Hall current in such a system has been detected through non-local measurements (Figure 7b). The origin of non-local signal in a graphene/hBN device can be considered as the combination of the VHE and a reverse VHE. The current applied (J) between any two contacts (Figure 7b), induces a valley current, $J_v$ in the perpendicular direction to J. The valley current creates an imbalance in valley populations, $\delta_\mu = \mu_K - \mu_{K'}$ between two distant contacts ($\mu_{K,K'}$ are the chemical potentials in the valleys $K$ and K'). As a result, a voltage (E) response is generated due to the reverse VHE. In this scenario

$$J_v = \sigma_{xy}^v E, \qquad E = \frac{\sigma_{xy}^v \rho_{xx}}{2e} \nabla \delta\mu \qquad (24)$$

Where, $\sigma_{xy}^{v}$ is the valley Hall conductivity and $\rho_{xy}$ is the longitudinal resistivity. The charge-neutral valley current can persist over extended distances in the absence of intervalley scattering and generate non-local electrical signals.[189-192] The VHE-induced non-local resistance ($R_{NL}$) can be expressed by the relation,[70]

$$R_{NL} = (\omega/2\xi)(\sigma_{xy}^{v})^2 \rho_{xx}^3 \exp(-L/\xi) \qquad (25)$$

where L is the distance between the current path and voltage probes, w is the device width and ξ is a constant. Sui et al. described a method of artifact-free measurement of $R_{NL}$.[31]

A non-zero value of $R_{NL}$ is observed if there is valley polarization. The graphene/hBN aligned devices exhibit large sharp peaks of $R_{NL}$ at the main and hole-side non-local points (NPs). However, no non-local signal has been detected in the (graphene/hBN) nonaligned devices where valley polarization is absent (Figure 7c) and in the gapless bilayer graphene (BLG) in which inversion symmetry is present (Figure 7d). The BLG under a transverse electric field exhibits a giant non-local response because of the topological transport of the valley pseudospin (Figure 3d-g).[31] Actually, in BLG, the intrinsic inversion symmetry is broken by the perpendicular electric field which resulted in valley polarization.[76, 77, 82]

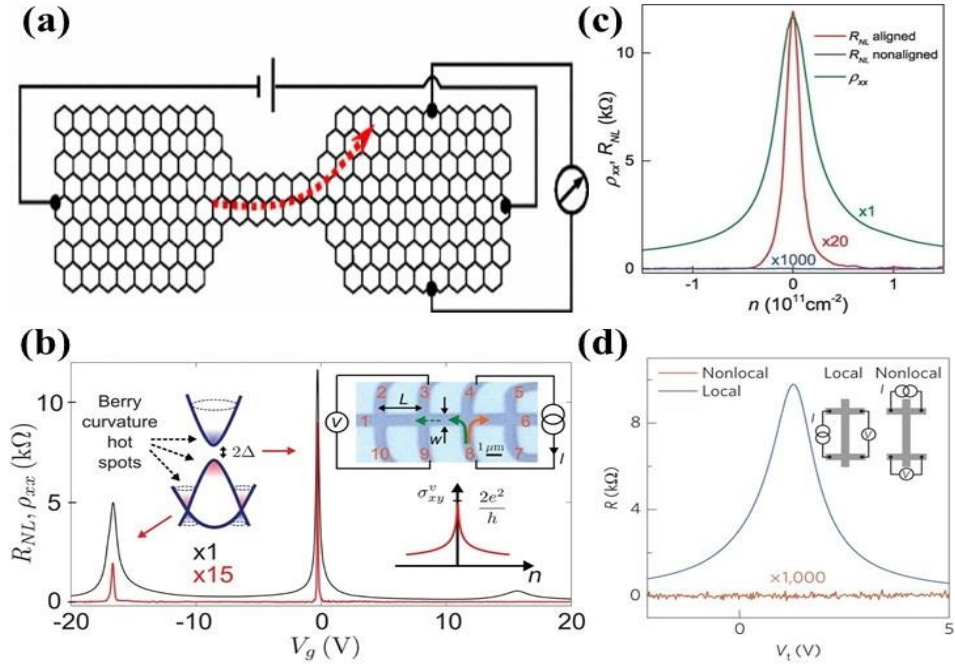

**Figure 7** (a) A transverse voltage across the sample due to valley polarization created by the valley filter.[26] (b) Non-local resistance in graphene superlattices (red curve) and longitudinal resistance (black curve) measured in graphene/hBN superlattices. Top right inset: Optical micrograph of graphene/hBN device used for the non-local measurement. Left inset: Illustration of band structure of graphene superlattices, showing Berry curvature hot spots. Bottom right inset: Modeled valley Hall conductivity for gapped Dirac fermions as a function of carrier density.[70] (c) Longitudinal and Hall resistivities ($\rho_{xx}$) and $R_{NL}$ near main non-local point in superlattice devices (green and red). $R_{NL}$ is zero for nonaligned device (blue).[70] (d) Variation of local (blue) and non-local (red) resistance with gate voltage in a BLG sample. Inset: Schematic representation of local and non-local measurement setup.[31]

The valley polarization in TMDs can be monitored by measuring the photoinduced anomalous Hall effect (AHE).[76] A Hall bar device of TMDC with a long Hall probe and a short conduction channel (Figure 8a) is used to measure the Hall effect. The reasons for such device structure are to produce photocurrent near the center of the device so that the Hall probe can efficiently pick up any Hall voltage, and to reduce the background photovoltage generated at the metal-semiconductor contacts of the Hall probe. A circularly polarized light is shined onto the Hall bar device to produce valley polarization in TMDs through breaking of time-reversal symmetry (Figure 8a). The Hall voltage is measured between the A and B contacts of the Hall probe (Figure 8a) by applying a source-drain voltage ($V_x$) across the short channel. A positive Hall voltage is observed for a monolayer $MoS_2$ device for a positive bias when the polarization of the incident light is modulated from right- to left (R-L)- handed (Figure 8b).[76] The sign of Hall voltages is reversed in the case of left- to right (L-R)- handed modulation of the polarization of the incident light. However, no Hall voltage is developed for linear (s-p) polarization. The photoinduced AHE in monolayer $MoS_2$ bears the signature of a net valley polarization under polarized light. On contrary, AHE is completely absent in bilayer $MoS_2$ (Figure 8b) suggesting no valley polarization in bilayer because of the unbroken inversion symmetry.[76] Wavelength-dependent mapping of the VHE in a transistor (made out of TMDs) suggested that the VHE in TMDs arises upon illumination with light having equal to the energy of excitons or trions (Figure 8c).[193] The VHE in TMDs is generated directly due to the trions. But excitons break at the interface of metal and semiconductors, generating free electrons and holes that result in a VHE. The intrinsic VHE has been observed without any extrinsic symmetry breaking in both monolayer and trilayer $MoS_2$ even at room temperature.[194] A very similar non-local signal due to VHE suggests that VHE is mainly contributed by the valence-band carriers in monolayer and trilayer $MoS_2$. The VDF has been identified even in multilayer $WSe_2$ through probing the valley Hall effect.[195] A current is generated in the presence of a drain-source voltage as a consequence of photoinduced VHE (Figure 8d). The valley polarization in multilayer $WSe_2$ was induced by breaking the spatial-inversion symmetry by applying a transverse electric field through an ionic liquid gate.

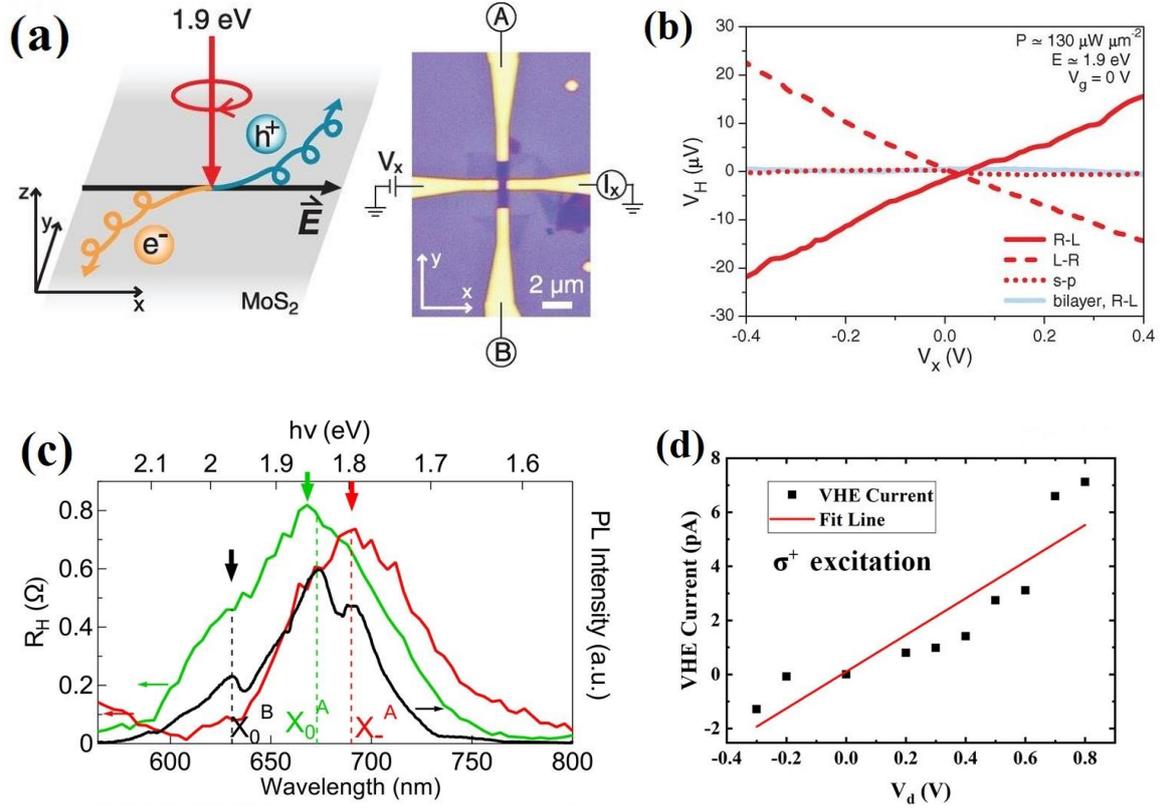

**Figure 8** (a) Schematics of a hall bar device (right) and representation of anomalous hall effect induced by valley polarisation (left).[76] (b) Dependence of the Hall voltage on the source-drain bias ($V_x$) of monolayer device under R-L (red solid line), L-R (red dashed line), S-P (red dot line) modulation, and bilayer device under R-L modulation (blue solid line).[76] (c) Spectral dependence of the Hall resistance in a bilayer 3R-$MoS_2$ device.[193] (d) Valley Hall current varies linearly with drain voltage $V_d$ in multilayer $WSe_2$.[195]

### 4.3 Valley Zeeman effect

Investigation of the resonance of exciton and other pseudo-states in the presence of an external magnetic field is an excellent way to study the immeasurable properties of semiconductor materials. In recent years, several experiments have been developed in the magneto-optical domain for the transition metal dichalcogenides.[103, 104, 137, 196, 197] One of such methods is the Valley Zeeman effect which is used to identify valley polarization in TMDs. In conventional Zeeman effect, a spin-dependent energy shift of the electronic energy levels takes place in the presence of an external magnetic field that lifts the degeneracy. Similarly, the valley pseudospin of an electron in TMDs can interact with external magnetic field producing Valley Zeeman effect. Valley Zeeman splitting and magnetic tuning of polarisation have been observed by polarisation resolved magneto reflectance and magneto-PL measurements.[197, 198]

Helicity resolved photoluminescence in an external magnetic field has been performed on the perpendicular plane of mechanically exfoliated TMDC layer.[197] To resolve the splitting between two valley excitons, the sample was both excited and detected with the same polarized light. The splitting can be determined by addressing one valley at one time and comparing the peak position of PL for different polarisations. The representative PL spectra are corresponding to the excitonic peaks at different values of the magnetic field, B, is shown in Figure 9(a). In the case of zero magnetic fields, the PL peak associated with each valley, i.e., $K^+$ (blue, $\sigma^+$) and $K^-$ (red, $\sigma^-$) is identical (middle). On the other hand, at high magnetic field (+7 T, top), both the components spectra were split with $\sigma^+$ at slightly lower energy than $\sigma^-$. In contrast, at low magnetic field (-7 T, bottom), $\sigma^+$ spectral component was shifted to the higher energy than $\sigma^-$ component (Figure 9(c)). The observed splitting is the consequence of the magnetic moment of the d- orbital of tungsten and the valley magnetic moment, which is the lattice contribution of the Berry curvature.[20] Figure 9(b) shows the Zeeman splitting of the bands due to each of these two contributions. The conduction and valance bands at zero (positive) magnetic field are denoted by dashed (solid) lines. Due to the spin splitting of ~0.4 eV in the valance band, the $K^+$ ($K^-$) valley has only spin-up(down) states. For the conduction band, the splitting is small with opposite signs in two valleys.[199, 200] The total Zeeman splitting of each band is determined by the sum of these three magnetic moment contributions. The Zeeman shift arising due to spin magnetic moment in conduction band is $\Delta_s = 2s_z \mu_B B$ where $\mu_B$ is the Bohr magneton (black arrow), whereas the atomic contribution in valance band is $\Delta_\alpha = 2\tau_z \mu_B B$ (purple arrow)), which contribute no shift in conduction band. The Zeeman shift arises due to the valley magnetic moment is given by $\Delta_v = \alpha_i \tau \mu_B B$ $\Delta_v = 2\alpha_i \tau_z \mu_B B$ (green arrow), where $\alpha_i$ is valley g-factor.

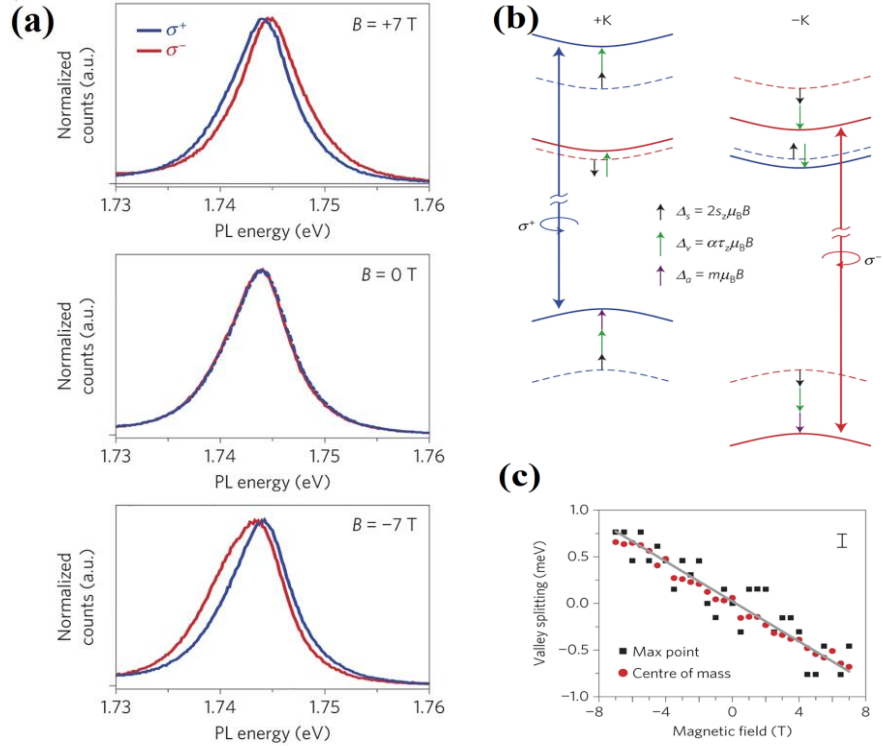

**Figure 9** (a) Energy level diagram for the three contributions to the Valley Zeeman splitting i.e., magnetic (black), atomic (purple), and valley (green) magnetic contributions. (b) Helicity resolved photoluminescence spectra of WSe$_2$ at different magnetic fields. (c) Valley Zeeman splitting as a function of magnetic field.[197]

Polarisation resolved magneto-PL and magneto-RC measurements for MLs of semiconducting MoTe$_2$ in magnetic field up to 29 T has been carried out by A. Arora and et al.[198] The σ⁺ (orange) and σ⁻ (blue) components of micro-reflectance contrast (μRC) spectra were measured for monolayer MoTe$_2$ exfoliated on Si/SiO$_2$ (Figure 10(a)) and sapphire substrates (Figure 10(b)) at different applied magnetic fields. The μRC spectrum can be obtained from the relation,[198]

$$C(\lambda) = [R(\lambda) - R_0(\lambda)]/ [R(\lambda) + R_0(\lambda)] \tag{26}$$

where, R(λ) and R₀(λ) are wavelength-dependent reflectance spectra of the TMDC layer and the bare substrate, respectively. Zeeman splitting can be clearly visible from X⁰$_A$ resonance, but the line shape of resonance remains unchanged with the variation of the magnetic field. The vale of Zeeman splittings for the A and B excitons can be used to find the g-factors. For ML MoTe₂ on sapphire substrate, the value of g is −4.7 ± 0.4 and −3.8 ± 0.6 for the A and B excitons, respectively. Polarisation resolved micro-PL spectra for monolayer MoTe₂ at various applied magnetic fields can be seen in Figure 10(c). A clear Zeeman splitting is visible in $X^{\pm}$ and $X_A^0$ transitions. The transition energies for both the σ⁺ and σ⁻ polarizations vary as a function of magnetic field (Figure 10(c)). The Zeeman splitting decreases with the increase of the magnetic field (Figure 10(e). The average value of g-factor estimated from PL measurements for neutral excitons (A and B) are $g_{X_A^0} = -4.7 \pm 0.4$ and $g_{X_B^0} = -3.8 \pm 0.6$ for the ML on sapphire substrate.[198] The valley polarisation is induced in TMDs under an external magnetic field. Although the degree of circular polarization is zero in the absence of magnetic field, it increases for neutral and charged exciton up to 78% and 36 % respectively, at a magnetic field of 29T (Figure 10(f)). Many groups have demonstrated vally polarization in TMDs through the Valley Zeeman effect.[198, 201-203]

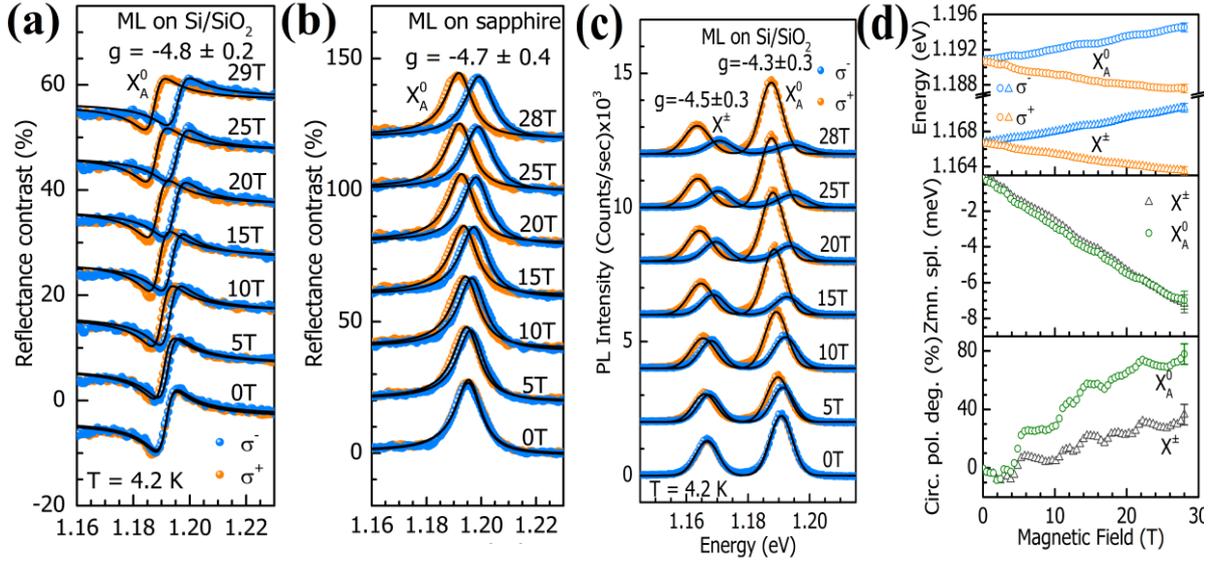

**Figure 10** (a) Polarisation resolved micro-reflectance contrast spectra of mechanical exfoliated monolayer MoTe$_2$ on Si/SiO$_2$ and (b) on sapphire substrates.[198] (c) Helicity resolved micro-PL spectra of neutral and charged excitons in monolayer MoTe$_2$ on Si/SiO$_2$ as a function of magnetic field (solid black lines are the fitted curves).[198] (d), (e), and (f) Zeeman spitted transition energies, valley Zeeman splitting, and degree of circular polarization as a function of magnetic field, respectively.[198]

## 5. Achieving and controlling the valley degree freedom in optoelectronic devices

The main challenge for developing valleytronic devices is to achieve the ability to generate and manipulate permanent valley polarization. In TMDCs, the optical excitonic transitions are dependent not only on spin DOF but also associated with valley DOF.[55] Therefore, it is possible to interconvert valley polarization with the optical polarization of emitted and absorbed photons in 2D TMDs. This means valley polarization can be carried from one system to another by photons. Such valleytronics devices can be realized by developing a valley-polarized optoelectronic emitter and absorber (a valley optical interconnect).

In a recent report, Zhang et al. have been demonstrated electrically switchable multilayer and monolayer WSe$_2$ p-n junctions which emit circularly polarised electroluminescence based on the VDF of the material.[204] They formed the junction using electrical double layer (EDL) gating with multilayer $WSe_2$. EDL gating provides a very large gate field due to accumulation of carries inside the 2D channel. There are two major difficulties with multilayer TMDCs, first, they have an indirect bandgap which is not preferable for optoelectronic device application, and second, restored inversion symmetry decreases the circular polarisation. EDL gating overcomes these difficulties as the large gate field produces more carrier densities in direct gap K and K' valleys and breaks inversion symmetry. Hence the multilayer ELD transistor operates in a similar way to monolayer TMDC transistor.[205] The electroluminescence (EL) spectra of the EDL transistor are circularly polarized and EL intensity increases with increasing gate voltage (Figure 11a). The degree of polarisation of the EL is comparable with the PL in monolayer TMDCs.[34, 185] The EL polarization is affected by the change of biasing of source and drain, i.e., when the device is operated in p-n mode (current flows from right to left), left circularly polarized emission dominates. In contrast, it decreases in n-p mode (Figure 11a). This behavior could be understood on the lights of variation in the overlapping of electron-hole distribution in K and K' valleys as the biasing is reversed (Figure 11b).

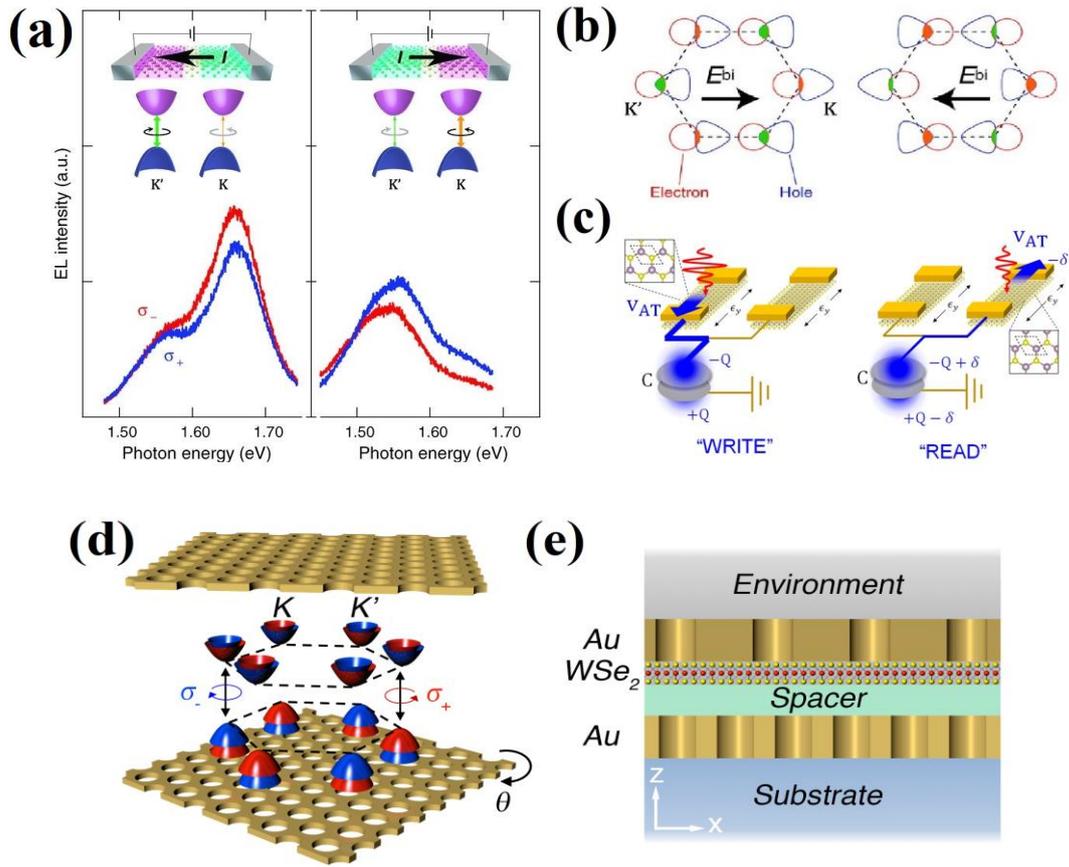

**Figure 11** (a) Top: schematic illustration of circularly polarised EL for two opposite current directions. Bottom: contributions of EL from K and K' valleys.[204] (b) Schematic representation of electron-hole distribution and overlapping (orange and green for K and K' valleys, respectively).[204] (c) Schematics of READ and WRITE operations in a single bit dynamical random-access memory (DRAM).[206] (d) Schematic of the WSe$_2$-MCM hybrid system. The monolayer WSe2is represented in reciprocal space, illustrating the band structure and the optical selection rule.[207] (e) Cross-sectional schematic showing the design of the WSe$_2$-MCM hybrid systems.[207]

The research team of Jae Dong Lee discovered the formation of valley magnetic domain (VMD) in TMDCs (MoS$_2$).[206] Because of VMD TMDCs can be used as a semiconducting device which is different from the existing devices that allow the current to flow only in one direction under the influence of external voltage. The control of electric current can be achieved through the VMD manipulation. The reversal of the electric-field direction leads to the VMD switching which can be exploited in monolayer TMDCs to realize transverse diode. A single bit dynamical random-access memory (DRAM) has been demonstrated using two transverse diodes (two strained $MoS_2$ ribbons) (Figure 11(c)).[206] In the write process, a far-infrared wave (FIR) falls on "WRITE" channel (the $MoS_2$ ribbon on the left) producing a transverse current to charge the capacitor, which records 1 to the bit (Figure 11c left). FIR is applied on "READ" channel (the $MoS_2$ ribbon on the right) producing a transverse current flow out from the capacitor to discharge the capacitor during the read process (Figure 11c, right). In such memory devices, read and write processes can be performed with THz speed.

The chiral valley Hall states have been generated in between the oppositely gated 2D graphene.[208, 209] Such states are called the kink states which enable several in situ transmission control mechanisms. The kink states of bilayer graphene can be exploited for the applications of waveguide, valve, and electron beam splitter.

The optical manipulation of the valley behaviors has been successfully achieved at cryogenic temperature. Still, the phonon-assisted intervalley scattering increases at high temperature, resulting in volatile valley states which reduce handedness of PL at room temperature. Valley-optical cavity hybrid systems can be used to maintain handedness of PL at room temperature.[210] However, control of valley properties at room temperature is still limited because of imperfect spatial and spectral overlap between excitons and optical cavities. Recently, Li et al. demonstrated valleytronic building blocks at room temperature similar to transistors in electronics.[211] They have generated long lived valley-polarized carriers in $MoS_2$ under linearly polarized infrared excitation using chiral nanocrescent plasmonic antennae. In fact, the field enhancement and near-field coupling properties of surface plasmon resonance have been widely exploited to modulate valley polarization of TMDCs. A recent review discussed the mechanisms of plasmonic modulation of valleytronic emission in TMDCs with surface plasmon polaritons (SPP) and localized surface plasmons (LSP).[212] Wu et al. reported a delicate way to manipulate quantum-information carriers – i.e., pairs of positive and negative charges confined at momentum valleys at room temperature in a monolayer $WSe_2$.[207] The new manipulation strategy is based on the strong light-matter interactions between the valley excitons and a plasmonic chiral metamaterial consists of two layers of periodic Au nanohole arrays with a controllable interlayer in-plane rotation angle (θ) (Fig 11(d, e)).

## 6. Summary and future scope

In this review, we have presented a detailed account of the origin of VDF in 2D materials having hexagonal lattice. There has been significant experimental work investigating valley polarization by creating charge carriers, excitons, and other quasi-particles in graphene and TMDCs. Unfortunately, these carriers of valley pseudospin are short-lived and rapidly jump between the valleys, which hinders the storage of information for a longer time. Though interlayer and defect-bound excitons have shown promise in realizing long-lived valley-selective excitons, a concerted effort will require to identify valley carriers of a long lifetime.

The dark excitons and trions could be excellent valley carriers as they possess a long lifetime and better valley stability than the bright excitons and trions. A dark exciton or trion can decay into a pair of phonon and photon with a particular handedness depending on the direction of the decay process. The handedness of the emitted photon could be used to read the valley indices of the dark states, but research in this direction is still in the early stages.

Despite the rapid progress in the field of generation and modulation of valley polarization, there remain significant challenges towards the practical development of valleytronic devices. One critical challenge is that almost all of the valley-dependent properties discovered till date are observed at cryogenic temperature. Further efforts are required to generate stable valley signature at room temperature and to develop methods for reading the valley indices for eventual applications of valleytronic technology.

**Acknowledgements**

AS and SKP acknowledge the financial support from Science and Engineering Research Board (SERB), Government of India (Grant No. CRG/2018/003045).

**Conflict of Interest**

The authors declare no conflict of interest.